\documentclass[preprint,amsmath,amssymb,showpacs,groupedaddress]{revtex4}
\usepackage{graphicx}
\usepackage{dcolumn}
\usepackage{bm}
\usepackage{epsfig}
\usepackage{amsmath}
\usepackage[letterpaper,left=25mm,right=25mm,top=30mm,bottom=20mm]{geometry}
\begin{document}
\draft
\title{Entanglement Creation and Storage in
Two Qubits Coupling to an Anisotropic Heisenberg Spin Chain}
\date{\today}
\author{Chunlei Zhang}
\author{Shiqun Zhu\footnote{Corresponding author, E-mail: szhu@suda.edu.cn
}}
\author{Jie Ren}

\affiliation{School of Physical Science and Technology, Suzhou
University, Suzhou, Jiangsu 215006, People's Republic of China}

\begin{abstract}
The time evolution of the entanglement of a pair of two spin qubits
is investigated when the two qubits simultaneously couple to an
environment of an anisotropic Heisenberg $XXZ$ spin chain. The
entanglement of the two spin qubits can be created and is a periodic
function of the time if the magnetic field is greater than a
critical value. If the two spin qubits are in the Bell state, the
entanglement can be stored with relatively large value even when the
magnetic field is large.
\end{abstract}

\pacs{03.67.Mn, 42.50.Dv, 03.65.Ud}

\maketitle

\section{Introduction}

Entangled quantum states are used mainly for quantum information
processing, such as quantum teleportation, quantum secret-code and
quantum computation \cite{Nielsen,Bennett,Murao}. Many
investigations showed that entanglement exists naturally in the spin
system when the temperature of the system is at zero
\cite{Vidal,Osborne,Osterloh,Amico}. In recent years, the study of
the dynamics of entanglement
\cite{Ciccarello,Konrad,Yu,Paz1,Paz2,Tiersch,Hamdouni,Wang} has
attracted much attention in the manipulation of quantum systems. The
dynamical properties and the time evolution of the entanglement in
different quantum systems were investigated. These systems included
mobile particle elastically-scattered by static spins
\cite{Ciccarello}, quantum mixed states \cite{Konrad,Yu}, two
oscillators coupled to the same environment \cite{Paz1,Paz2}, two
d-level systems \cite{Tiersch}, decoherence of a spin-$1/2$ particle
coupled to a spin bath in thermal equilibrium \cite{Hamdouni}, a
spin chain in driving the the decoherence of a coupled quantum
system \cite{Wang}, etc. Meanwhile, the effects of the environment
were taken into account. The excitation and quantum information
transfer was investigated between two external spins when they
coupled to a one-dimensional spin chain at different sites
\cite{Hartmann}. The entanglement induced by two external spins
could be used to signal the critical points when they were
simultaneously coupled to an environmental $XY$ spin chain
\cite{Yi,Yuan}. The decay of the Loschmidt echo was enhanced by the
quantum criticality of the surrounding Ising chain when an external
spin was coupled to the environment \cite{Quan}. When two external
spins coupled to a transverse field Ising chain, the induced
entanglement could be enhanced near quantum criticality and could be
used to detect the quantum phase transition \cite{Ai}, which
occurred in the many-body quantum systems \cite{Sachdev}. The
dynamical properties of the entanglement in a spin system need to be
further investigated when it is coupled to an external environment.

In this paper, the dynamics and the time evolution of the
entanglement of a pair of two qubits are investigated when the two
qubits simultaneously couple to an environment of an anisotropic
antiferromagnetic Heisenberg spin chain with magnetic field. In
Section II, the Hamiltonian of the system and the effective
Hamiltonian of the two qubits coupled to the environment are
presented. In Section III, the time evolution of the system is
analyzed for the simplest case of the environment. The entanglement
creation in the coupled pair of two external qubits is discussed in
Section IV. In Section V, the storage of the entanglement in the
coupled pair of two external qubits is investigated. A discussion
concludes the paper.

\section{Hamiltonian of the System}

When two external spin qubits are coupled with the environment of a
one-dimensional spin chain, the Hamiltonian of this system can be
written as
\begin{equation}
\label{eq1} H =H_0+H_I.
\end{equation}
where $H_{0}$ is the Hamiltonian of the environment. If the
environment is an anisotropic Heisenberg $XXZ$ spin chain, one has
\begin{equation}\label{eq2}
H_{0}=J\sum_{i=1}^{N}(\sigma^x_i\sigma^x_{i+1}+\sigma^y_i\sigma^y_{i+1}+\Delta_i\sigma^z_i\sigma^z_{i+1})+B\sum_{i=1}^{N}\sigma^z_i,
\end{equation}
were $\sigma^{\alpha}_i(\alpha=x, y, z)$ are Pauli operators, $N$ is
the number of the spin chain, $J$ is the coupling coefficient
between the spins, $B$ is the magnetic field along the z-axis with
the anisotropy $\Delta_i=\Delta$ $\in$ (0,1). In Eq. (1), $H_{I}$ is
the interaction Hamiltonian between the two external spin qubits and
the environment and can be written as,
\begin{equation}\label{eq3}
H_{I}=J_{p}\sum_{i=1}^{N}(\sigma_a\sigma_i + \sigma_b\sigma_i).
\end{equation}
where $\sigma_a$ and $\sigma_b$ are the Pauli operators of the
qubits $a$ and $b$, $J_{p}$ is the coupling coefficient between the
external spin qubits ($a$ and $b$) and the Heisenberg spin chain. In
order to facilitate the calculation, the coupling coefficients are
chosen as $J=1$ and $J_p=0.2$ in this paper. That is, the
environment is represented by the antiferromagnetic Heisenberg $XXZ$
spin model. The schematic diagram of the system is shown in Fig. 1.
The two qubits are symmetrically located at the two sides of the
spin chain.

Fr\"{o}hlich transformation \cite{Ai,Frohlich} can be used to solve
the problem of induced effective interaction between two qubits
through the medium of the Heisenberg spin chain. The environment of
the antiferromagnetic Heisenberg spin chain has non-degenerate
ground state $|\psi_{0}\rangle$ with ground state energy $E_{0}$.
According to the standard canonical transformation
\cite{Ai,Frohlich,Ferreira}, the effective Hamiltonian of the
external spin qubits can be written as

\begin{equation}\label{eq4}
H^{ab}_{eff}=\sum_{j=1}^{k}\frac{\langle\psi_{0}|H_iP_{j}H_i|\psi_{0}\rangle}{E_{j}-E_{0}},
\end{equation}
where the projector is $P_j=|\psi_{j}\rangle\langle\psi_{j}|$ and
$|\psi_{j}\rangle (j=1,2,...k)$ is the time dependent excited state
with energy $E_j$. After some straightforward calculations, the
effective Hamiltonian can be reduced to

\begin{equation}\label{eq5}
H^{ab}_{eff}=-\sum_{j}2J_pJ_p\sum_{\alpha,\beta}\Re(m_{\alpha}n^{\ast}_{\beta})\sigma^{\alpha}_{a}\sigma^{\beta}_{b}+
\sum_{\alpha,\beta}\frac{J^2_{p}}{4}(|m_{\alpha}|^{2}+|n_{\beta}|^{2}),
\end{equation}
where the parameters are
$m_{\alpha}=\frac{\langle\psi_{0}|s^{\alpha}_{m}|\psi_{j}\rangle}{\sqrt{E_{k}-E_{0}}},
n_{\beta}=\frac{\langle\psi_{0}|s^{\beta}_{n}|\psi_{j}\rangle}{\sqrt{E_{k}-E_{0}}},
s^{\alpha,\beta}=\frac{1}{2}\sigma^{\alpha,\beta}$, and
$\Re(m_{\alpha}n^{\ast}_{\alpha})$ means the real part of the
product $(m_{\alpha}n^{\ast}_{\alpha})$ with $\alpha,\beta=x,y,z$.
When the eigenstate $|\psi_{j}\rangle$ and the corresponding
eigenvalue $E_{j}$ of $H_0$ are obtained, the effective Hamiltonian
$H^{ab}_{eff}$ can be easily calculated.

\section{Analysis of Time evolution}

In order to describe the time evolution of the entanglement of
two-qubit system, the concurrence is used as a measure of the
entanglement. The concurrence is defined as \cite{Hill,Wootters}
\begin{equation}\label{eq6}
C = \max \{{\lambda_1 - \lambda_2 - \lambda_3 - \lambda_4 ,0}\},
\end{equation}
where the $\lambda_i (i=1, 2, 3, 4)$ are the square roots of the
eigenvalues of the density matrix $\varrho_{ab}$. The density matrix
$\varrho_{ab}$ is given by

\begin{equation}\label{eq7}
\varrho_{ab}=\rho_{12}(\sigma_1^y \otimes \sigma_2^y)\rho_{12}^\ast
(\sigma_1^y \otimes \sigma_2^y).
\end{equation}

The ground state of the environment of the Heisenberg spin chain can
be chosen as $|\phi_{0}\rangle$ while that of the two external spin
qubits $a$ and $b$ can be chosen as $|01\rangle$. Under the
influence of the environment, the two external spin qubits have an
initial state as follows

\begin{equation}\label{8}
|\psi_{0}\rangle=|\phi_{0}\rangle\otimes|01\rangle_{ab}.
\end{equation}
The time evolution of the state is

\begin{equation}\label{9}
|\psi(t)\rangle=\exp(-iH^{ab}_{eff}t)|\psi_{0}\rangle_{ab},
\end{equation}
with the density matrix
$\varrho_{ab}=|\psi(t)\rangle\langle\psi(t)|$.

The reduced density matrix $\varrho_{ab}(t)$ can be written as
\begin{equation}\label{10}
\varrho_{ab}(t)=\left(
\begin{array}{cccc}
u(t) & 0    & 0    & 0 \\
0    & w_1(t) & y(t) & 0 \\
0    & y^\ast(t) & w_2(t) & 0 \\
0    & 0    & 0    & v(t) \\
\end{array}
\right)
\end{equation}
in the standard basis $\{|00\rangle, |01\rangle, |10\rangle,
|11\rangle\}$. The corresponding concurrence $C(t)$ of the two
external spin qubits can be calculated from the reduced density
matrix $\varrho_{ab}(t)$ and given by

\begin{equation}\label{11}
C(t)=2\max\{|y(t)|-\sqrt{u(t)v(t)}, 0\}.
\end{equation}

\section{Entanglement Creation}

For the simplest case of $N=2$ in the anisotropic Heisenberg $XXZ$
spin chain, the eigenenergies and eigenstates of the system are
$E_1=\Delta-2B, E_2=\Delta+2B, E_3=-\Delta+2, E_4=-\Delta-2$ and
$|\varphi_{1}\rangle=|11\rangle, |\varphi_{2}\rangle=|00\rangle,
|\varphi_{3}\rangle=\frac{\sqrt{2}}{2}(|01\rangle+|10\rangle),
|\varphi_{4}\rangle=\frac{\sqrt{2}}{2}(-|01\rangle+|10\rangle)$
respectively. When $B-\Delta>1$, the ground state is
$|\phi_0\rangle=|\varphi_1\rangle$. Then the effective Hamiltonian
$H^{ab}_{eff}$ can be written as

\begin{equation}\label{eq12}
H^{ab}_{eff}=g\left(
\begin{array}{cccc}
1 & 0 & 0 & 0 \\
0 & 1 & 2 & 0 \\
0 & 2 & 1 & 0 \\
0 & 0 & 0 & 1 \\
\end{array}
\right),
\end{equation}
where the parameter $g$ is given by
$g=J_{p}^{2}\frac{\Delta-B}{(B-\Delta+1)(B-\Delta-1)}$. The matrix
elements $u(t), w_1(t), y(t), w_2(t)$ and $v(t)$ in Eq. (10) are
given by $u(t)=0,
w_1(t)=\frac{1}{2}+\frac{1}{4}e^{-i4gt}+\frac{1}{4}e^{i4gt},
y(t)=-\frac{1}{4}e^{-i4gt}+\frac{1}{4}e^{i4gt},
w_2(t)=\frac{1}{2}-\frac{1}{4}e^{-i4gt}-\frac{1}{4}e^{i4gt},
v(t)=0$. When $B-\Delta<1$, the effective Hamiltonian
$H^{ab}_{eff}=0$. There is no entanglement between spin qubits $a$
and $b$. When $B-\Delta=1$, the ground state energy equals the
excited state energy, i. e., $E_1=E_4$. The energies of the two
states are crossed at this point. Since the two states are
degenerate, Eq. (4) is not valid to calculate the effective
Hamiltonian $H^{ab}_{eff}$ when $B-\Delta=1$. That is, there is a
critical value of the magnetic field $B_C$. The value of $B_C$ is
given by $B_C=1+\Delta$. If $B<B_C$, the concurrence $C(t)$ is zero.
That is, there is no entanglement when $B<B_C$. If $B>B_C$, the
entanglement appears. That is, the entanglement can be created when
$B>B_C$. Then the concurrence $C(t)$ can be given by

\begin{equation}\label{eq13}
C(t)=\left\{
     \begin{array}{ll}
     0, & (B-\Delta<1); \\
     |\sin(4gt)|, & (B-\Delta>1). \\
     \end{array}
     \right.
\end{equation}

The concurrence $C(t)$ as a function of the time $t$ is plotted in
Fig. 2 when the magnetic field $B$ and the anisotropy $\Delta$ are
varied. The values of the anisotropy are $\Delta=0.2, 0.4$ and $0.6$
with $B>B_C$ in Figs. 2(a), 2(b) and 2(c) respectively. From Fig. 2,
it is seen that the concurrence $C(t)$ is a periodic function of
time $t$. It almost oscillates between the maximum value of one and
the minimum value of zero. The period decreases as the magnetic
field $B$ increases.

The anisotropic antiferromagnetic Heisenberg $XXZ$ model was used to
investigate the order-to-disorder transition of the material
$Cs_2CoCl_4$ \cite{Kenzelmann}. For the material $Cs_2CoCl_4$, the
anisotropy is $\Delta=0.25$. When the number of spins in the
environment of the Heisenberg $XXZ$ chain is greater than two, there
is no approximate analytic solution of $H^{ab}_{eff}$ and $C(t)$. To
calculate $C(t)$, the numerical computation needs to be performed.
In Fig. 3, the concurrence $C(t)$ is plotted as a function of time
$t$ when the spin numbers in the environment are $N=4, 6$ and $8$.
From Fig. 3, it is seen that the concurrence $C(t)$ is a periodic
function of $t$ with two different kinds of periods. Both periods
decrease as the spin number $N$ in the environment increase. There
is a critical value $B_C$ of the magnetic field. When $B<B_C$, the
concurrence $C(t)$ oscillates following the large period. The period
decreases slightly as $B$ increases. While $B>B_C$, $C(t)$
oscillates following the small period. The period increases as $B$
increases. The concurrence $C(t)$ can be approximately given by
\begin{equation}\label{eq14}
C(t)\sim\left\{
     \begin{array}{ll}
     |sin[g(N)\sqrt{N/(N+1)}t]|, & (B<B_C); \\
     |sin[g(N)Nt]|, & (B>B_C). \\
     \end{array}
     \right.
\end{equation}
Where $g(N)$ is a function of the spin number $N$ in the
environment. Though there is no analytic expression of the critical
field $B_C$, it can be numerical calculated. The critical field
$B_C$ is plotted in Fig. 3(d) as a function of $1/N$. From Fig.
3(d), it is seen that $B_C$ decreases linearly as $1/N$ decreases.
In the thermodynamic limit of $N\rightarrow \infty$, $B_C$ tends to
zero. The regime for larger period of oscillation disappears.

\section{Entanglement storage}

The concurrence $C(t)$ is plotted as a function of magnetic field
$B$ and time $t$ in Fig. 4 when the initial state of the two
external spin qubits $a$ and $b$ is in the Bell state
$\frac{1}{\sqrt{2}}(|01\rangle+|10\rangle)$. The anisotropy is
chosen as $\Delta=0.25$ \cite{Kenzelmann}. The spin numbers in the
environment of the anisotropic antiferromagnetic Heisenberg chain
are $N=2, 4, 6$, and $8$. From Fig. 4, it is seen that the
concurrence $C(t)$ is a oscillation function of time $t$. The
oscillation period decreases as $B$ increases. Obviously, the
concurrence $C(t)$ is divided into several regions by different
critical values of the magnetic field $B_C$. The red circles in Fig.
4 show the critical values of $B_C$. At the critical point $B_C$,
the energies of the ground and excited states are crossing and the
states are degenerate. For $N=2$, there are two parts in $C(t)$
divided by one $B_C$. For $B<B_C$, $C(t)$ is almost a constant of
$C(t)=1.0$. For $B>B_C$, $C(t)$ oscillates with a small period [cf.
Fig. 4(a)]. For $N=4$, there are three parts divided by two
different values of $B_C$ [(marked by two red red circles in Fig.
4(b)]. In the first part, $C(t)$ is very close to $1.0$. It
oscillates with quite small amplitude. In the second part, $C(t)$
oscillates with small period. In the third part, $C(t)$ oscillates
with even smaller period [cf. Fig. 4(b)]. When $N=6$ and $8$ in
Figs. 4(c) and 4(d), similar phenomena occurs. Obviously, the
concurrence $C(t)$ is divided into $(N/2+1)$ parts by $N/2$ critical
values of $B_C$. The energy is crossing at the critical values of
$B_C$ in the ground state as well as in excited states.The
concurrence $C(t)$ is jumping as the state is switched from an
entangled state to another. In the thermodynamic limit, the
continuous energy level crossings occur \cite{Son}. The first part
of concurrence $C(t)$ disappears. Other parts of $C(t)$ tends to
smooth and continuous. In Fig. 3(d), only the first critical value
of $B_C$ as a function of spin number $1/N$ is plotted. From Fig. 4,
it is also clear that the entanglement $C(t)$ can keep large value
even for relatively large magnetic field $B$. If the initial state
of the two external spin qubits is the Bell state
$\frac{1}{\sqrt{2}}(|00\rangle+|11\rangle)$, similar results as that
shown in Fig. 4 are obtained.

\section{Discussion}

The time evolution of the entanglement of two external spin qubits
is investigated when they are coupled to the environment of an
anisotropic antiferromagnetic Heisenberg $XXZ$ spin chain with
magnetic field. The approximate form of the effective Hamiltonian is
derived. The concurrence is used as a measure of the entanglement.
When there are two spins in the environment, there is no
entanglement between two external spin qubits when the magnetic
field is smaller than a critical value. When the magnetic field is
greater than the critical value, the entanglement can be created and
is a periodic function of the time. The entanglement almost
oscillates between one and zero. The oscillation period decrease as
the anisotropy and the magnetic field increase. There are $N/2+1$
parts in the entanglement divided by $N/2$ values of critical
magnetic fields. The first critical magnetic field tends to zero
when the spin number in the environment tends to infinity. When the
initial state of the two external spin qubits is in one of the Bell
state, the entanglement can be stored. Though there are different
regimes in the entanglement, the entanglement always keeps quite
large value when it oscillates with increasing number of spins in
the environment.

\vspace{0.5 cm}

{\bf Acknowledgments}

It is a pleasure to thank Xiang Hao and Tao Song for their many
helpful discussions. The financial support from the National Natural
Science Foundation of China (Grant No. 10774108) is gratefully
acknowledged.

\newpage

\begin{center}{\Large \bf Figure Captions}\end{center}

\textbf{Fig. 1}

The schematic diagram of two external spin qubits symmetrically
coupled to the environment of an anisotropic Heisenberg spin chain.

\textbf{Fig. 2}

The concurrence $C(t)$ is plotted as a functions of the time $t$ for
$N=2$ when the magnetic field $B$ and the anisotropy $\Delta=$ are
varied with $B>B_C$. (a). $\Delta=0.2$. (b). $\Delta=0.4$. (c).
$\Delta=0.8$.

\textbf{Fig. 3}

The concurrence $C(t)$ is plotted as a function of the magnetic
field $B$ and the time $t$ for different spin numbers $N$ of the
environment in (a), (b), and (c). The anisotropy is $\Delta=0.25$.
(a). $N=4$. (b). $N=6$. (c). $N=8$. (d). The critical field $B_C$ is
plotted as a function of $1/N$.

\textbf{Fig. 4}

The concurrence $C$ is plotted as a function of the magnetic field
$B$ and the time $t$ for different spin numbers in the environment.
The anisotropy is $\Delta=0.25$ and the Bell state
$\frac{1}{\sqrt{2}}(|01>+|10\rangle)$ is chosen. (a). $N=2$. (b).
$N=4$. (c). $N=6$. (d). $N=8$.

\newpage

\centerline{\epsfig{file=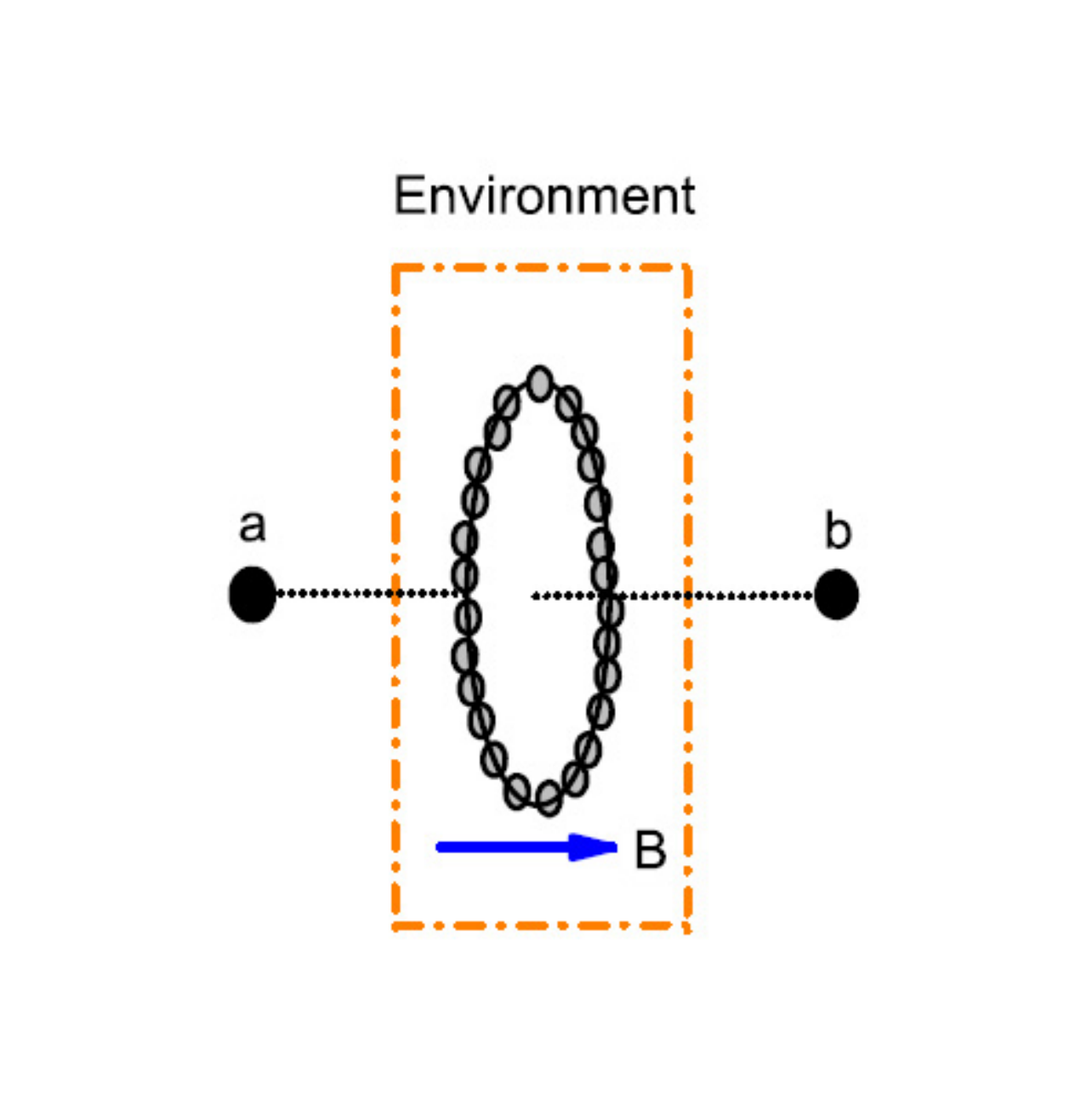,width=250pt}} \centerline{\bf
\large Fig. 1}
\newpage
\centerline{\epsfig{file=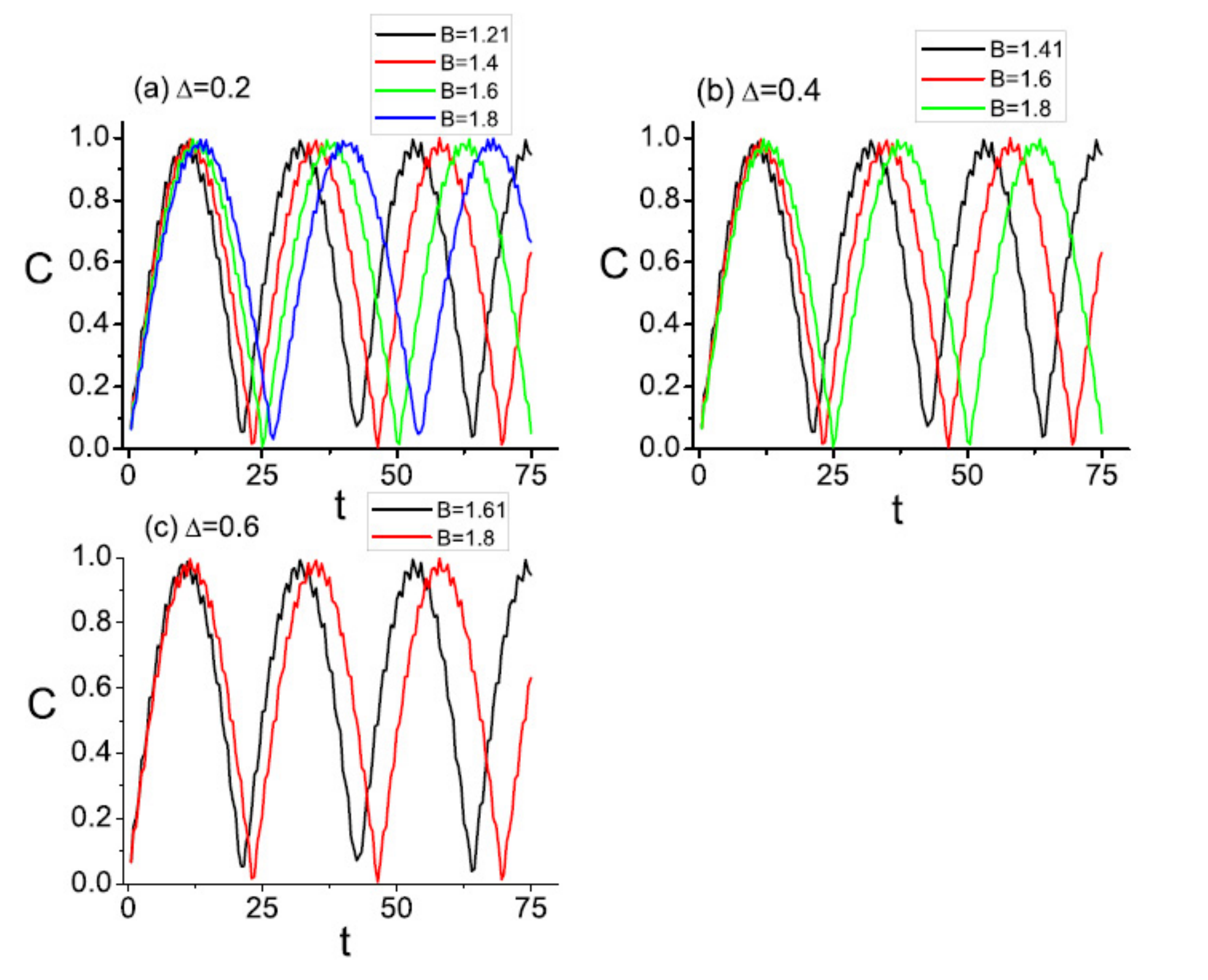,width=400pt}} \centerline{\bf
\large Fig. 2}
\newpage
\centerline{\epsfig{file=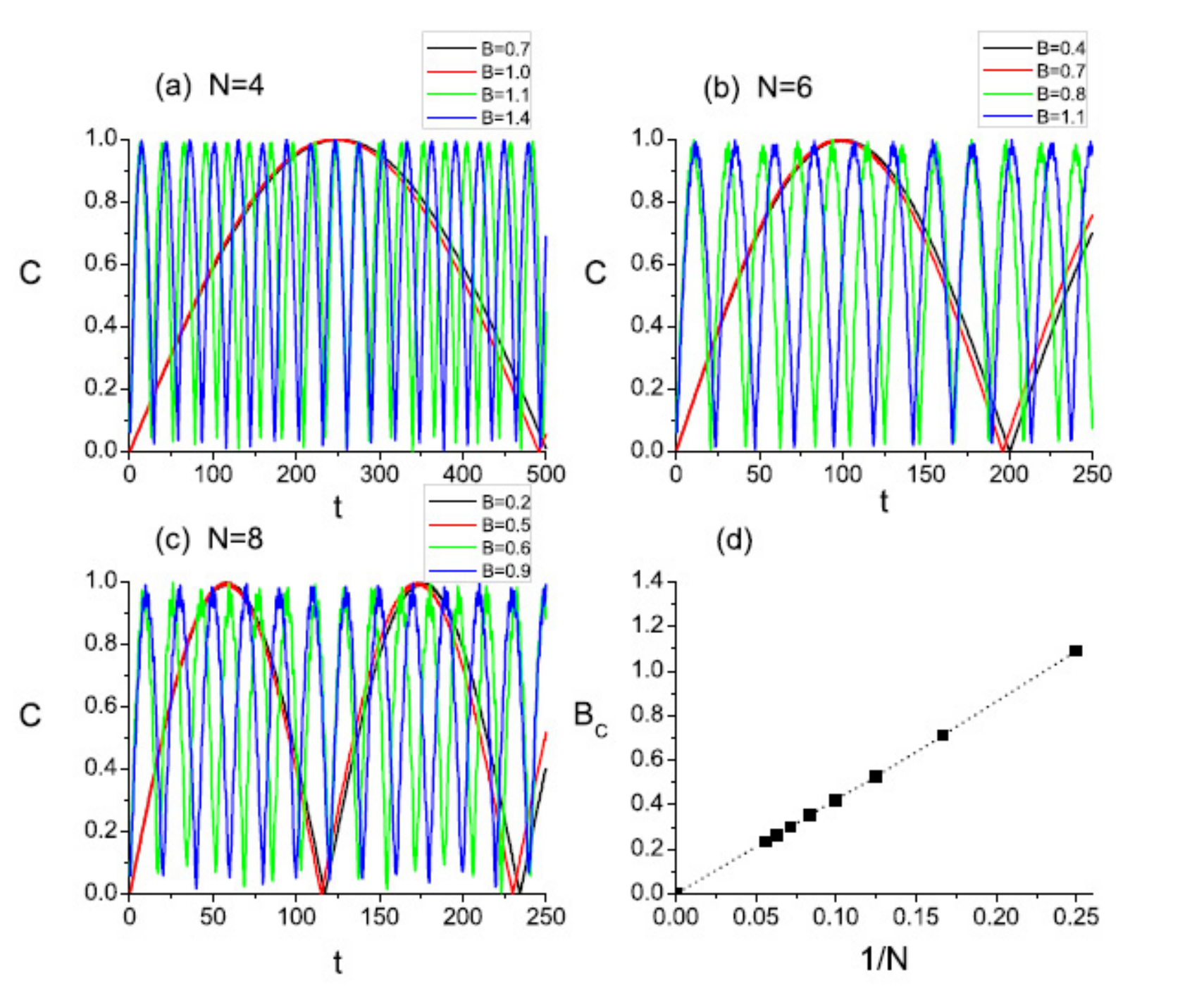,width=400pt}} \centerline{\bf
\large Fig. 3}
\newpage
\centerline{\epsfig{file=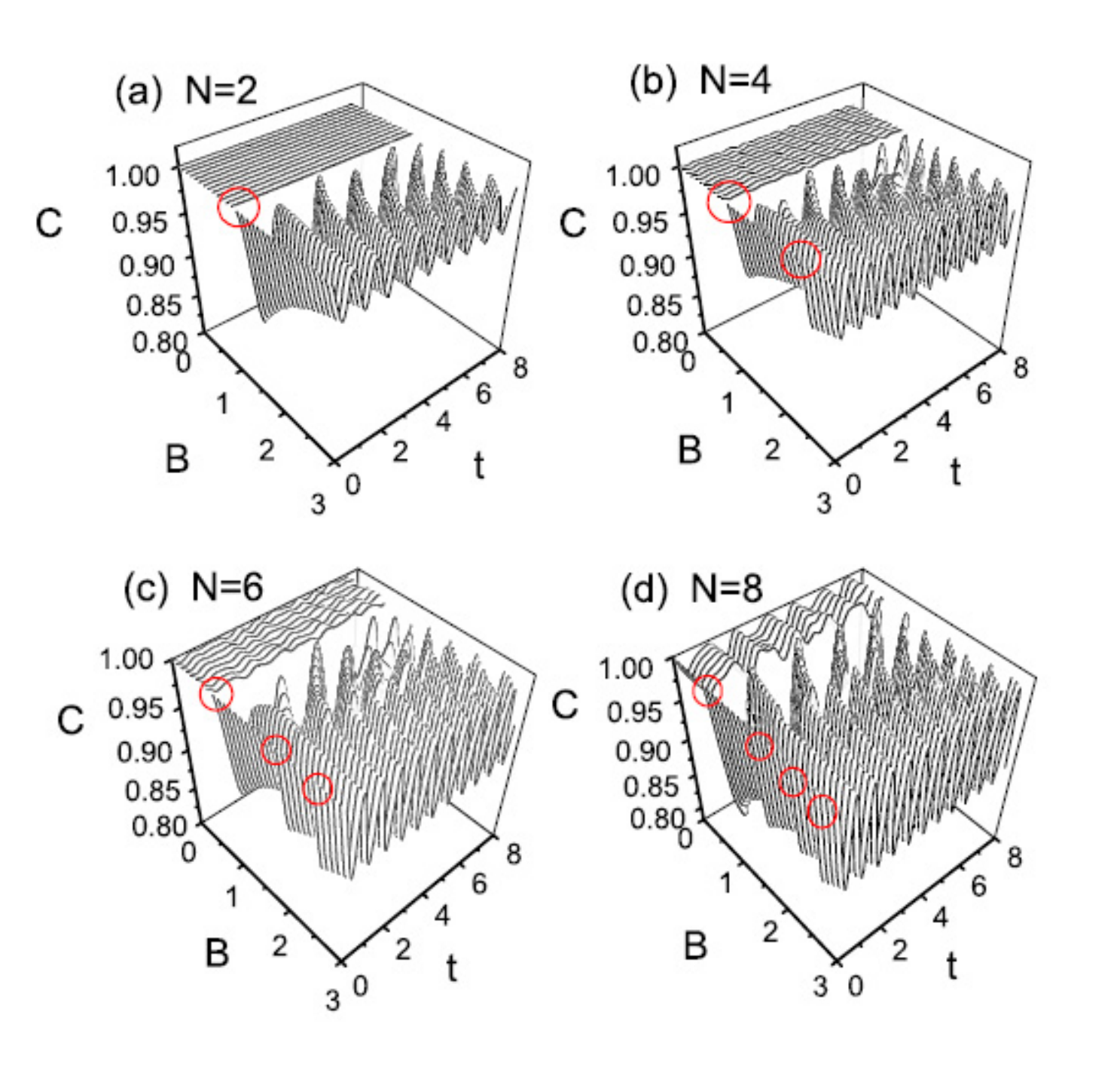,width=400pt}} \centerline{\bf
\large Fig. 4}

\end{document}